\documentclass[conference]{IEEEtran}
\IEEEoverridecommandlockouts

\usepackage{cite}
\usepackage{amsmath,amssymb,amsfonts}
\usepackage{graphicx}
\usepackage{textcomp}
\usepackage{xcolor}
\usepackage{adjustbox}
\usepackage{flushend}

\usepackage{algorithm}
\usepackage{algpseudocode}
\usepackage{subfig}

\usepackage{float}
\usepackage[left=1.62cm,right=1.62cm,top=1.9cm]{geometry}

\def\BibTeX{{\rm B\kern-.05em{\sc i\kern-.025em b}\kern-.08em
    T\kern-.1667em\lower.7ex\hbox{E}\kern-.125emX}}

\begin{document}

\title{NR Cell Identity-based Handover Decision-making Algorithm for High-speed Scenario within Dual Connectivity\\
\thanks{This work was supported in part by the Japan Society for the Promotion of Science (JSPS) under KAKENHI Grant JP22H03585.}
}

\author{
\IEEEauthorblockN{Zhiyi Zhu\textsuperscript{1,*}, 
 Eiji Takimoto\textsuperscript{2}, 
 Patrick Finnerty\textsuperscript{1},
 Junjun Zheng\textsuperscript{3}, 
 Shoma Suzuki\textsuperscript{1},  
 Chikara Ohta\textsuperscript{1}}
\\
\IEEEauthorblockA{\textsuperscript{1}\textit{Graduate School of System Informatics, Kobe University, Kobe, Japan}}
\IEEEauthorblockA{\textsuperscript{2}\textit{Information Technology Center, Nara Women's University, Nara, Japan}}
\IEEEauthorblockA{\textsuperscript{3}\textit{Graduate School of Advanced Science and Engineering, Hiroshima University, Hiroshima, Japan}}
\IEEEauthorblockA{* Corresponding author email: syu@port.kobe-u.ac.jp}
}

\maketitle

\begin{abstract}
The dense deployment of 5G heterogeneous networks (HetNets) has improved network capacity. 
However, it also brings frequent and unnecessary handover challenges to high-speed mobile user equipment (UE), resulting in unstable communication and degraded quality of service. 
Traditional handovers ignore the type of target next-generation Node B (gNB), resulting in high-speed UEs being able to be handed over to any gNB.
This paper proposes a NR cell identity (NCI)-based handover decision-making algorithm (HDMA) to address this issue. 
The proposed HDMA identifies the type of the target gNB (macro/small/mmWave gNB) using the gNB identity (ID) within the NCI to improve the handover decision-making strategy. 
The proposed HDMA aims to improve the communication stability of high-speed mobile UE by enabling high-speed UEs to identify the target gNB type during the HDMA using the gNB ID. 
Simulation results show that the proposed HDMA outperforms other HDMAs in enhanced connection stability.
\end{abstract}


\section{Introduction}
5G, the fifth generation of cellular networks, faces the challenge of maintaining seamless connectivity and channel stability in high-speed scenarios~\cite{9813555}~\cite{10549894}. 
The current handover decision-making algorithm (HDMA) only considers channel measurement indicators, including reference signal received power (RSRP), reference signal received quality (RSRQ), or signal-to-interference-plus-noise (SINR~\cite{9767158}. 
Consequently, the UE can trigger a handover to any gNB if these conditions are met.
User equipment (UE) moving at high speeds will remain in a small gNodeB (gNB) for a short time, and frequent handovers by the current HDMA are inevitable, resulting in degraded or even interrupted communication quality, which has a significant impact on the UE experience~\cite{10.1145/3636534.3690680}~\cite{CHEN202578}.

The dual connectivity (DC) scheme mitigates the problem of UE connection loss during hard handover to some extent by enabling the UE to connect to two base stations simultaneously and increase the throughput of the UE~\cite{7959177}.
However, the presence of various types of gNBs in the heterogeneous ultra-dense network, including millimeter waves (mmWave) and small gNBs within sub-6 GHz and due to the severe path loss of millimeter waves (mmWave), the high mobility of UEs, and the numerous obstacles present in urban environments, frequent intermittent interruptions of UE wireless links occur~\cite {10145424}.
These problem of degraded communication quality caused by frequent handovers of highly mobile UEs still needs to be addressed and optimized~\cite{8108450}~\cite{https://doi.org/10.1049/cmu2.12814}.

\section{Related work}
\label{sec:rel-work}

Recently, there has been a lot of research on high-speed mobility management and handover optimization.
Some of the work focuses on using user mobility information to optimize handover decisions.
In~\cite{8650068}, fuzzy logic control is used to dynamically optimize the handover trigger timing based on UE speed and the cell size of a gNB, thereby reducing handover failures and ping-pong handovers.
In~\cite{9954717}, a HDMA is proposed for UE speed and service type that prevents the UE speed between 15 km/h and 30 km/h from handover to small cells by considering the UE service type and speed to maintain the downlink throughput.
In~\cite{8622167}, an adaptive handover scheme is proposed to adjust the threshold of the absolute value of RSRP to adjust the handover trigger conditions according to the UE speed.
In~\cite{10279306}, a HDMA based on multiple parameters, including received signal strength (RSS), equivalent RSS, handover margin (HOM), UE speed, quality of service (distinguishing between real-time services), and available bandwidth, is proposed. 
This HDMA aims to encourage high-speed users to either stay within or hand over to the macro base station to prevent frequent handovers.

In addition, machine learning and reinforcement learning techniques have also been widely used to optimize 5G handovers.
In~\cite{electronics12194131}, it employs a data-driven approach, utilizing a multilayer perceptron model to adaptively adjust handover control parameters such as HOM and Time-to-Trigger (TTT), while considering numerous factors, including UE speed, RSRP, SINR, and load. 
This approach has proven effective in achieving a significant reduction in handover failure rates, handover times, and ping-pong handover rates.
In~\cite{10463334}, a convolutional neural network (CNN)-long short-term memory (LSTM) deep learning model is employed to predict UE trajectories, enabling the selection of the target gNB in advance. 
This approach significantly reduces the number of unnecessary handovers and enhances service continuity.
In~\cite{KWONG2024122871}, it proposes an adaptive user-centric virtual cell handover decision oriented towards connectivity, which uses reinforcement learning with indicators such as RSRP and SINR to ensure the seamless connection of the UE to the virtual cell, thereby maintaining the stability of the communication channel for high-speed UEs.
In~\cite{7959177}, it proposes a deep reinforcement learning algorithm based on deep deterministic policy gradient (DDPG) to optimize the handover parameters autonomously. 
The thresholds, such as HOME, can be dynamically adjusted according to the environment, thereby enhancing the robustness of mobility at different speeds while maintaining high throughput and low latency. 
However, these learning-based solutions are predicated on a large amount of training data and offline optimization and are still insufficient for real-time response to network changes. 
Furthermore, these approaches do not use cell-level information.

Moreover, the communication environment for high-speed UE handover to mmWave cells is unstable due to high loss for mmWave gNBs after blocking obstacles.
In~\cite{riaz2025handover}, an intelligent handover mechanism is proposed that can dynamically fine-tune parameters such as HOM and TTT based on the real-time SINR to achieve a fast response to the drastic fluctuations of mmWave channels.
In~\cite{9120763}, it is designed to minimize handover interruption time by incorporating backhaul delay considerations during handover signaling exchange. 
Additionally, it employs a transmission control protocol (TCP) proxy to ensure lossless data transmission following handover.

The existing research on optimizing handover decisions in 5G high-speed environments heavily relies on instantaneous channel measurement indicators and parameter tuning. 
However, existing research does not adequately address the gNB type information carried by the NR cell identity (NCI). 
As a result, the current handover decision mechanism lacks knowledge of the characteristics of the target cell when determining whether to trigger a handover. 
In other words, neither the gNB nor the UE knows whether the target base station is a macro cell, a sub-6GHz small gNB, or an mmWave gNB. 
This oversight can lead to frequent handovers, causing an unstable communication environment.

To address this challenge, this study proposes, for the first time, a NCI-based mechanism for handover optimization in high-speed scenarios: NCI is used to determine the type of candidate gNB to avoid frequent handovers that lead to degraded communication quality.

Therefore, our contribution can be summarized as follows:
\begin{itemize}
    \item We propose a semantic extension of the NCI, in which the 22 bits gNB ID field is subdivided into a 2 bits gNB type ID and a 20 bits gNB ID.
    \item A NCI-based handover decision-making algorithm is proposed to enhance the stability of communication in high-speed scenarios.
    \item The simulation results show the robustness of the proposed HDMA in reducing the number of handovers and improving communication stability.
\end{itemize}

\begin{figure}[t]
\centering
\includegraphics[scale=0.45]{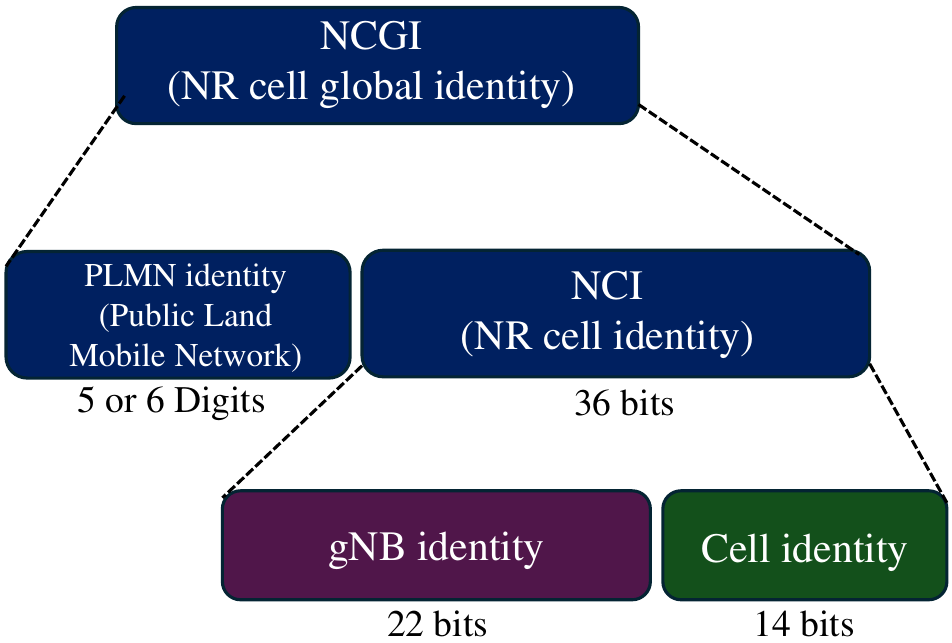} 
\caption{The composition of NR cell global identity}
\label{fig:nci_expl}
\end{figure}

The rest of this paper is organized as follows, 
First, we introduce the System model in Section~\ref{sec:sys-model}. 
Then, we introduce the proposed HDMA in Section~\ref{sec:prop-hdma}, followed by our evaluation results in Section~\ref{sec:eva}. 
Finally, we conclude the whole paper in Section~\ref{sec:concl}.

\section{System Model}
\label{sec:sys-model}
\subsection{Dual connectivity}
In 5G, dual connectivity refers to using multiple carriers supported by these gNBs for transmission and reception. 
This enables the UE to connect to both the master node (MN) and the secondary node (SN) simultaneously, thereby achieving broadband. 
The MN has a high output power and a low-frequency band. 
Thus, it has a large coverage area and can reliably exchange control signals with the UE; the SN has a low output power and a high-frequency band, so it has a short coverage area and can provide high data rates to the connected UE. 
Due to the MN's wide coverage, the UE always connects to the MN, and handover within the DC scenario is also referred to as a secondary node handover. 
In contrast to the hard handover scenario, in the DC scenario, when the UE experiences an SN handover, its data transmission is not completely interrupted, and the UE will communicate with the MN. 
Once the SN handover is complete, the UE will transmit data to the new SN. 
The MN channel is more stable than the SN channel, so if the UE's data transmission with the SN is interrupted due to handover or other reasons, the MN will immediately back up the data~\cite{9120763}.

\subsection{NR cell identity}

The NR cell global identity (NCGI) is composed of two components: the Public land mobile network (PLMN) identity of the cell and the NCI of the gNB, as illustrated in Fig.~\ref{fig:nci_expl}. 

Each gNB has a unique identity (gNB ID) that is used to distinguish different gNBs within the PLMN. 
Each gNB may have multiple cells, and each cell corresponds to a cell identity (cell ID).
The gNB ID and the cell ID, when combined, form the 36 bits NCGI, with the gNB ID configurable from 22 to 32 bits, with a minimum of 22 bits and a maximum of 32 bits.
The cell ID occupies 4 to 14 bits, with a minimum of 4 bits and a maximum of 14 bits. 
By default, the 22 bits gNB ID space is suitable for most operators' requirements. 

According to the 3GPP standard, no bits or specific codes are reserved in the 36 bits of the NCI to identify the cell type~\cite{3gpp.38.413}.
Therefore, it is necessary to consider what coding characteristics correspond to a specific gNB type through the current NCI, such as macro base station, small gNB, or mmWave gNB .

\begin{figure}[t]
\centering
\includegraphics[scale=0.45]{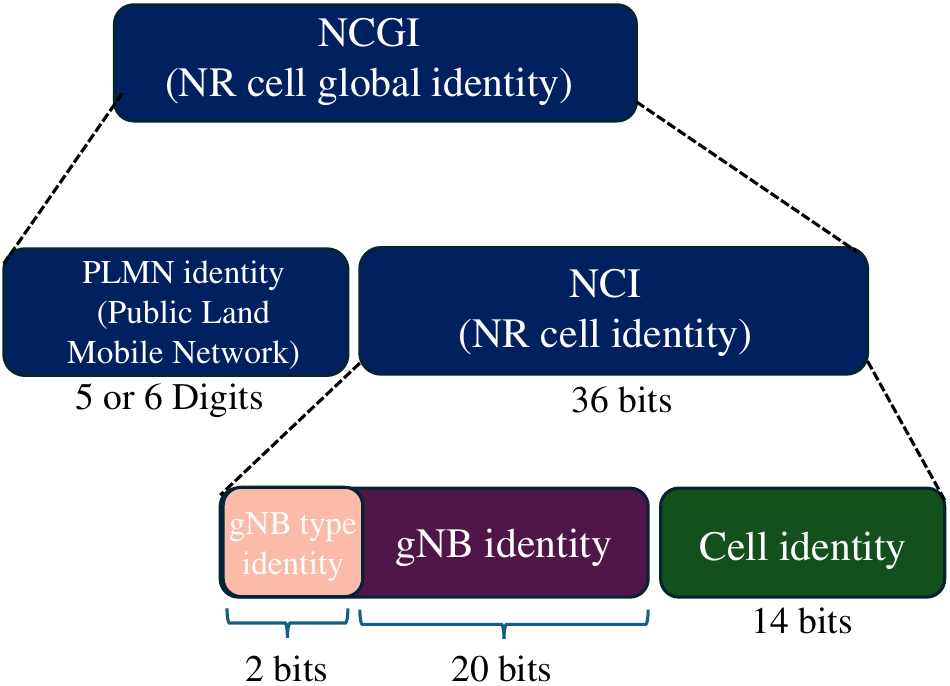} 
\caption{The proposed semantic extension of the NCI with gNB type identity}
\label{fig:prop_nci}
\end{figure}

In our semantic extension, we propose to split this 22 bits gNB ID into two subfields: the upper 2 bits, termed the 'gNB type ID', are reserved for encoding the gNB type (such as macro base station, sub-6GHz small, or mmWave gNBs), while the remaining 20 bits, termed the gNB ID, provide the unique identity for the gNB within its type. 
The detail is shown in Fig.~\ref{fig:prop_nci}.
The 14 bits cell ID does not change. 
This semantic extension has the advantage that the upper 2 bits within the gNB ID are used to represent the cell type, allowing the gNB type to be identified by the gNB type ID without the need for additional signaling.

\begin{algorithm}[t]
\caption{NCI-based handover decision algorithm}
\label{alg:nci-based}
\begin{algorithmic}[1]
\Require UE \textit {i}, speed \(v_{i}\), speed threshold \(T_v\), gNB list \(L\) with gNB type IDs, RSRP measurements, HOM \textit {H}, TTT \textit {ttt}
\If{$v_{i} \geq T_v$}
    \State \(C \gets \{\}\) \Comment{Initialize target gNB candidate list}
    \For{each gNB in \(L\)}
        \State \(tb \gets \text{Extract High Bits} \ b_i\) \Comment{\(b_i\) denotes the gNB type ID}
        \If{\(tb = \texttt{"01"}\) \textbf{or} \(tb = \texttt{"00"}\)}
            \State \(C \gets C \cup \{\text{gNB with \(tb\)}\}\)
        \ElsIf{\(tb = \texttt{"10"}\)}
            \State \textbf{continue} \Comment{Skip mmWave gNB}
        \EndIf
    \EndFor
    \State \(B \gets \text{Find best RSRP} \ C\) \Comment{Select the target gNB with the best RSRP}
    \If{\(\text{target gNB} \ RSRP > \text{current gNB} \ RSRP + H \ \text{within} \ \textit{ttt}\)}
         \State \text{Execute UE \textit {i} handover to \textit {B}}
    \Else
      \State Cancel UE \textit {i} handover
    \EndIf
\Else
     \State \text{Execute A3RSRP HDMA}
\EndIf
\end{algorithmic}
\end{algorithm}

\section{Proposed handover decision-making algorithm}
\label{sec:prop-hdma}

UE moving at high speed on a macrocell may only be able to obtain a low throughput rate due to its long distance from the base station and high load. 
However, a small cell in the surrounding area can improve the instantaneous signal quality and throughput even if the service time is short. 
While frequent handovers of high-speed UEs can degrade UE communication quality, small gNBs in the Sub-6 GHz band have better coverage and anti-obstacle capabilities than mmWave gNBs, making their links more stable in high-speed scenarios and providing more reliable signal strength.
However, in a 5G dense small gNBs scenario, if only instantaneous channel metrics such as RSRP, RSRQ, and SINR are used during HDMA, all potential gNBs will be included in the candidate target gNBs, resulting in frequent handovers, and communication degradation cannot be entirely avoided.
Therefore, it is considered to introduce NCI into HDMA only so that the UE can recognize the target gNB type during HDMA execution and select the target base station that matches its characteristics.
The proposed HDMA determination process is outlined in Algorithm~\ref{alg:nci-based}.

Specifically, the UE in a DC scenario with the presence of macro base station, sub-6 GHz small gNBs, and mmWave gNBs identifies that the target gNB is a sub-6 GHz small cell by the high two-bit coding within the NCI and includes it in the target gNB; the UE identifies that the target base station is a mmWave gNB by the high two-bit coding within the NCI and does not include it in the target gNB, to achieve a more stable communication environment. 

The details of the gNB ID semantic extension are shown in the following,
\begin{itemize}
    \item Macro base station type ID $\gets$ $00$
    \item Sub-6GHz gNB type ID $\gets$ $01$
    \item mmWave gNB type ID $\gets$ $10$
\end{itemize}

\section{Evaulation}
\label{sec:eva}

\begin{figure}[t]
\centering
\includegraphics[scale=0.45]{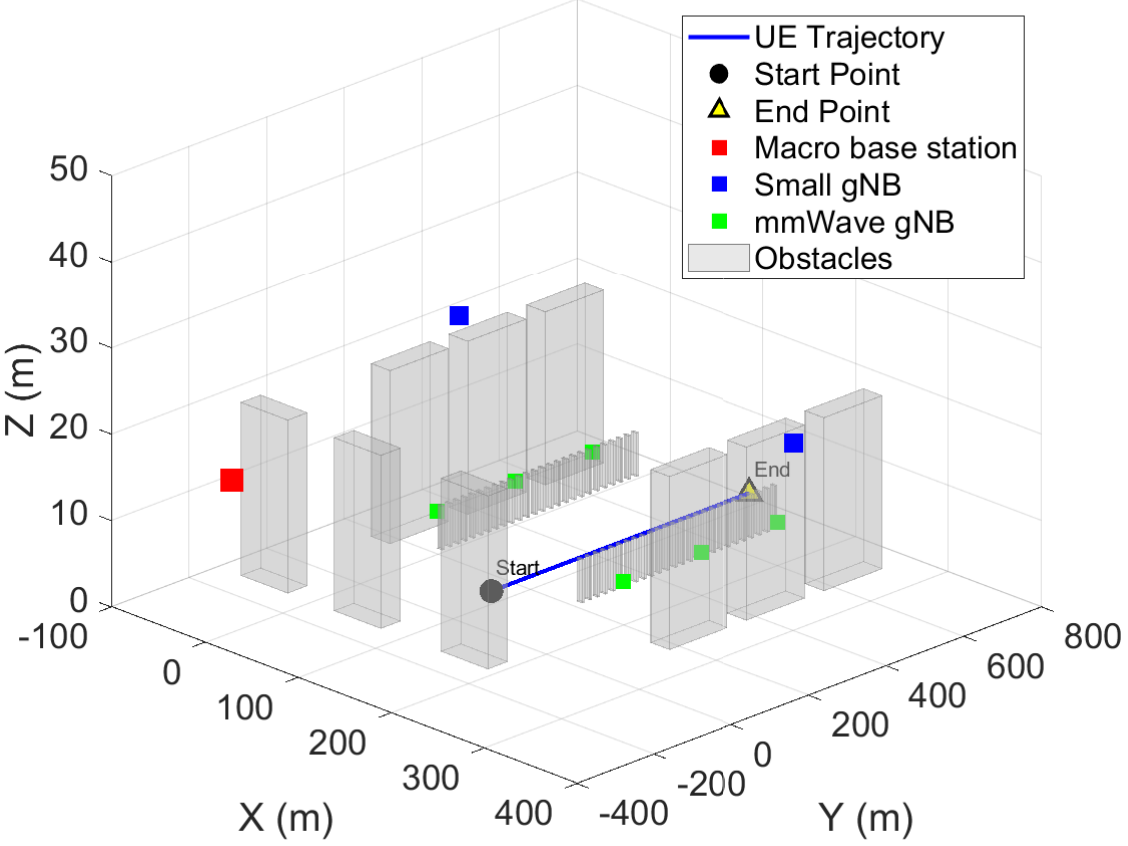} 
\caption{Simulation environment}
\label{fig:sim-env}
\end{figure}

As shown in Fig.~\ref{fig:sim-env}, we consider a 5G dual connectivity scenario that includes a macro base station within 2.1 GHz, two sub-6GHz small gNBs, and six mmWave gNBs, as a city scenario with multiple obstacles using MATLAB 5G Toolbox. 
This dual connectivity scenario is referenced from~\cite{9120763}, and we added two sub-6 GHz small cells to that.

\begin{table}[t]
\caption{Simulation parameters setting}
\label{tab:sim_params}
\centering
\renewcommand{\arraystretch}{1.2}
\begin{tabular}{|l|l|}
\hline
\textbf{Parameter} & \textbf{Value} \\
\hline
Simulation time & 40 s \\
Macrocell base station & 2.1 GHz \\
Small gNB & 3.5 GHz \\
mmWave gNB & 28 GHz \\
UE speed & 60 km/h \\
Speed threshold $T_{v}$ & 30 km/h \\
Handover margin & 3 dB \\
Time-to-Trigger & 200 ms \\
Packet sending interval & 0.001 s \\
\hline
\end{tabular}
\end{table}

Specifically, the macro base station is the MN in the dual-link scenario, and the small base station and millimeter wave base station are the SN in the dual-link scenario. 
Initially, the UE travels in a straight line along the Y-axis from the position $(100, 100, 1.5)$ at 60 km/h. 
The SN handover of the UE is triggered when the UE crosses the city.
The detail of the simulation setting is shown in Table~\ref{tab:sim_params}.

Additionally, to show the performance of the proposed HDMA, we compare it with the A3RSRP HDMA commonly used in 5G and the UE speed-based HDMA in reference~\cite{8650068} and~\cite{8622167}. 
The handover trigger condition based on A3RSRP HDMA is that the RSRP of the target gNB is greater than a threshold of the current base station RSRP within a short period of time; the handover trigger condition based on UE speed HDMA in referenc~\cite{8650068} and~\cite{8622167} are that when the UE movement is greater than a threshold, the UE data transmission is fixed to the macro base station to ensure the seamless communication.

\begin{figure}[t]
\centering
\includegraphics[width=0.7\linewidth]{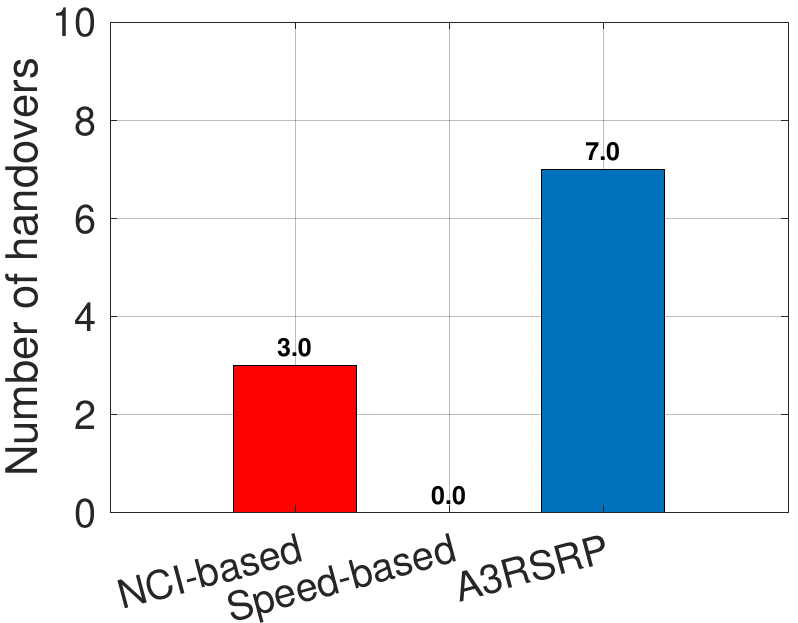}
\caption{Number of handovers.}
\label{fig:num-ho}
\end{figure}

\begin{figure}[t]
\centering
\includegraphics[width=0.85\linewidth]{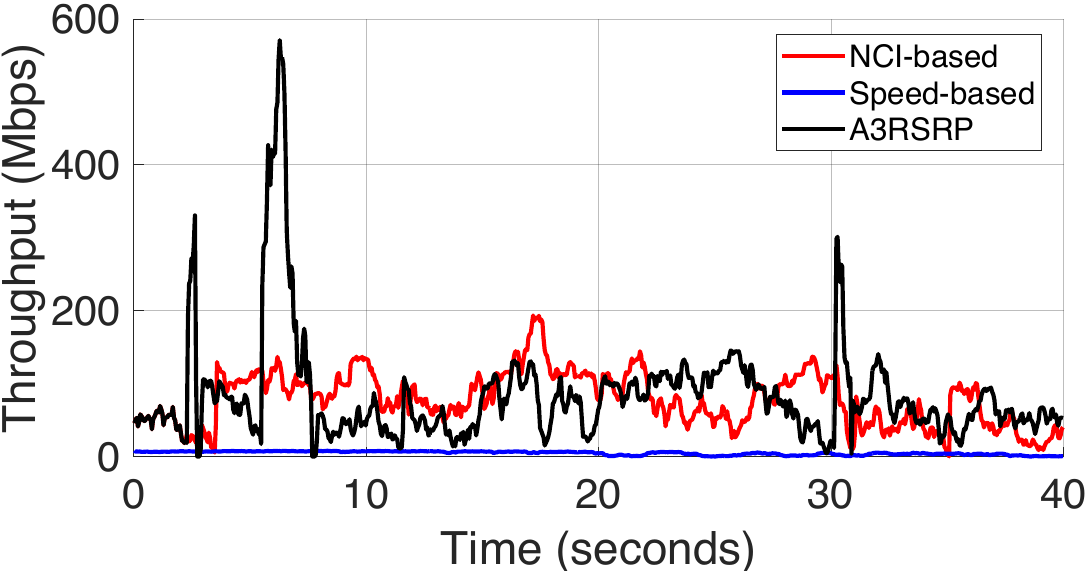}
\caption{Throughput vs time.}
\label{fig:throughput-time}
\end{figure}

\subsection{Handover numbers and throughput}

We first simulated and evaluated the number of handover triggers.
As illustrated in Fig.~\ref{fig:num-ho}, the number of handovers triggered by the three algorithms is shown. 
The results are NCI-based, Speed-based and A3RSRP HDMA from left to right, respectively.

As the results indicate, the number of handovers triggered by UE speed-based HDMA is 0 because the UE's movement speed exceeds the set speed threshold of 30 km/h. 
Consequently, the UE will not hand over to small cells or millimeter wave cells. 
The proposed NCI-based HDMA has a lower handover number than the A3RSRP HDMA. 
This is because the proposed NCI-based HDMA considers the NCI of the target gNB and UE movement speed, which enables avoidance of handover to the mmWave gNB and thereby reduces the number of handovers.

As illustrated in Fig.~\ref{fig:throughput-time}, although the handover number of HDMA based on UE movement speed is 0, the throughput is the lowest of the three, indicating that the communication environment in the macro cell is characterized by strong interference and high load, resulting in a low actual UE throughput.
Conversely, although the number of handovers for the proposed HDMA and A3RSRP HDMA is notably higher than that for UE-speed-based HDMA, the throughput performance of the former significantly surpasses that of the latter.

However, since A3RSRP HDMA does not identify the target gNB type, it hands over to the millimeter wave cell, resulting in an immediate increase in UE throughput and SINR, but also a significant degradation in communication due to obstruction.

\begin{figure}[t]
\centering
\includegraphics[scale=0.47]{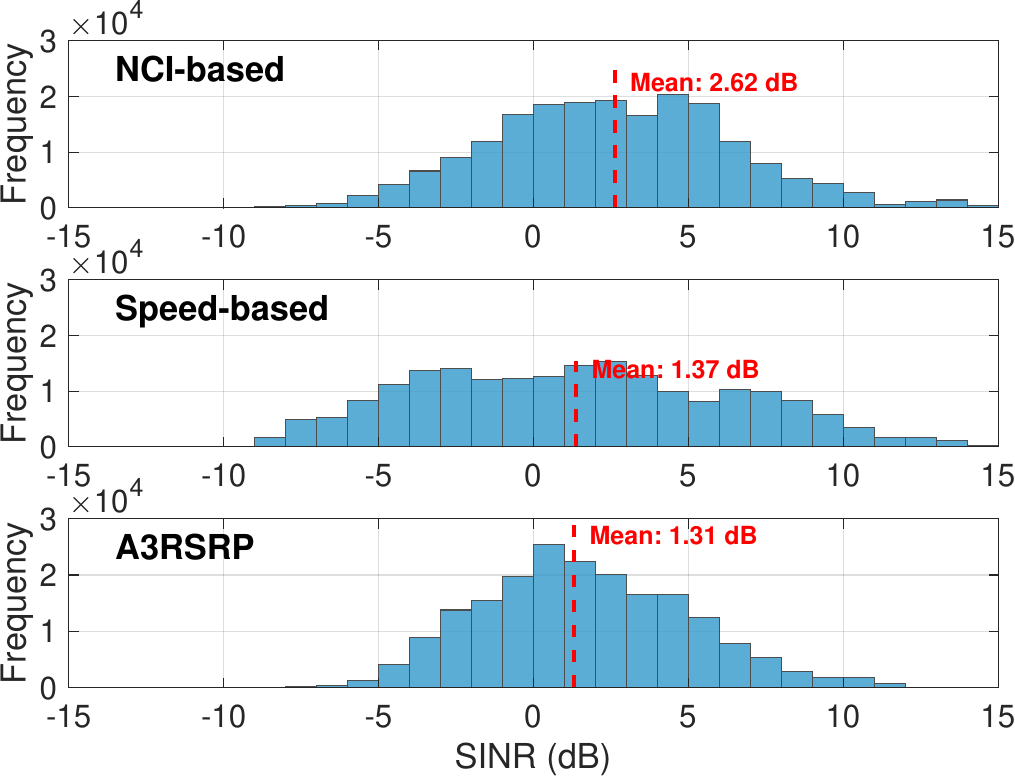} 
\caption{Comparison of the SINR data distribution}
\label{fig:sinr-hist}
\end{figure}

\begin{figure}[t]
\centering
\includegraphics[scale=0.45]{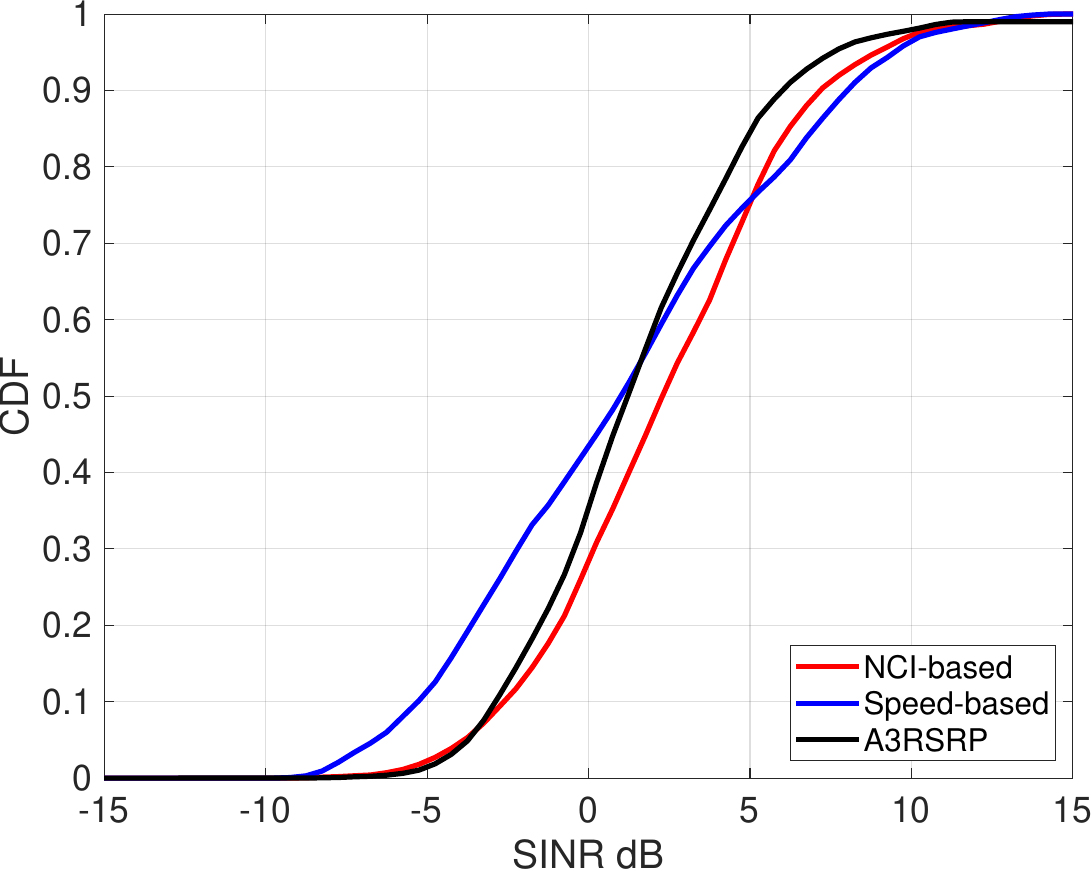} 
\caption{Comparison of the SINR CDF}
\label{fig:sinr-cdf}
\end{figure}

\subsection{Communication stability}

Fig.~\ref{fig:sinr-hist} shows an evaluation of the stability of the three algorithms, where the three sub-figures represent the histograms of the SINR data distribution of the three algorithms during the simulation time.
The top sub-figure in Fig.~\ref{fig:sinr-hist} shows the proposed NCI-based HDMA, the middle sub-figure shows the speed-based HDMA, and the bottom sub-figure shows the A3RSRP HDMA.

The speed-based HDMA does not hand over to other small cells in this DC simulation scenario, and the UEs always communicate with the macro base station.
Consequently, the SINR value remains consistent due to the lack of frequent handovers, and the average SINR is slightly higher than that of A3RSRP HDMA.
Conversely, the proposed NCI-based HDMA uses the gNB type ID to filter out millimeter-wave cells by identifying the target cell for handover in advance, and only sub-6 GHz small cells are used as the target cells for handover.
Consequently, the proposed NCI-based HDMA has a higher average SINR than the speed-based and A3RSRP HDMAs, and the SINR values occur more frequently in the 0 to 5 dB range.

Fig.~\ref{fig:sinr-cdf} shows the cumulative distribution function (CDF) of the SINR for the three HDMAs.
The proposed NCI-based HDMA prevents handovers to mmWave cells by identifying the target cell type in advance via the gNB type ID. Therefore, the SINR of the UE fluctuates more stably than with A3RSRP and speed-based HDMA.
Fig.~\ref{fig:sinr-hist} and Fig.~\ref{fig:sinr-cdf} demonstrate that the proposed NCI-based HDMA offers a more stable communication environment for the UE compared with speed-based and A3RSRP HDMA.

\section{Conclusion}
\label{sec:concl}

In this paper, we propose an extension of the NCI coding scheme based on the current NCI structure to help UEs identify the type of target base station during HDMA. This extension aims to provide a more stable communication environment for high-speed mobile UEs in dense 5G networks with multiple cellular types.
Through a comparative analysis of the proposed NCI-based HDMA with A3RSRP-based HDMA and speed-based HDMA, it has been demonstrated that the former can ensure higher throughput while reducing the number of handovers and exhibiting improved performance in terms of communication stability.

In future work, we will consider the scenario where multiple cells exist in a gNB, to test and optimize the performance of our algorithm, and consider the scenario where there are a large number of UEs with different mobility speeds, and ways to improve the performance of the algorithm by combining machine learning approaches.

\section{Acknowledgment}

This work was supported by JSPS KAKENHI Grant Number JP22H03585.

\bibliographystyle{IEEEtran}
\bibliography{ref}

\end{document}